\documentclass[sigconf]{acmart}

\usepackage{booktabs}
\usepackage[ruled,vlined,linesnumbered]{algorithm2e}
\usepackage{bbold}
\usepackage{setspace}

\usepackage{graphicx}
\usepackage{subcaption}

\newcommand{\todo}[1]{\textcolor{red}{#1}}

\newcommand{\sz}[1]{\textcolor{blue}{#1}}

\newcommand\captionshrink{\vspace*{-0.75\baselineskip}}

\newtheoremstyle{mydef}
{2ex}
{2ex}
{\itshape}
{}
{\scshape}
{: }
{0.5em}
{}
\theoremstyle{mydef}
\newtheorem{mydef}{Definition}
\begin{document}

\copyrightyear{2018} 
\acmYear{2018} 
\setcopyright{acmcopyright}
\acmConference[SIGIR '18]{The 41st International ACM SIGIR Conference on Research and Development in Information Retrieval}{July 8--12, 2018}{Ann Arbor, MI, USA}
\acmBooktitle{SIGIR '18: The 41st International ACM SIGIR Conference on Research and Development in Information Retrieval, July 8--12, 2018, Ann Arbor, MI, USA}
\acmPrice{15.00}
\acmDOI{10.1145/3209978.3209988}
\acmISBN{978-1-4503-5657-2/18/07}

\title{On-the-fly Table Generation}
\author{Shuo Zhang}
\affiliation{%
  \institution{University of Stavanger}
}
\email{shuo.zhang@uis.no}

\author{Krisztian Balog}
\affiliation{%
  \institution{University of Stavanger}
}
\email{krisztian.balog@uis.no}

\author{}

\begin{abstract}
Many information needs revolve around entities, which would be better answered by summarizing results in a tabular format, rather than presenting them as a ranked list.  Unlike previous work, which is limited to retrieving existing tables, we aim to answer queries by automatically compiling a table in response to a query.  We introduce and address the task of on-the-fly table generation: given a query, generate a relational table that contains relevant entities (as rows) along with their key properties (as columns).  This problem is decomposed into three specific subtasks: (i) core column entity ranking, (ii) schema determination, and (iii) value lookup.  We employ a feature-based approach for entity ranking and schema determination, combining deep semantic features with task-specific signals.  We further show that these two subtasks are not independent of each other and can assist each other in an iterative manner.  For value lookup, we combine information from existing tables and a knowledge base.  Using two sets of entity-oriented queries, we evaluate our approach both on the component level and on the end-to-end table generation task.
\end{abstract}



\maketitle

\vspace*{-0.5\baselineskip}
\section{Introduction}

\begin{figure}[t]
   \centering
   \includegraphics[width=0.50\textwidth]{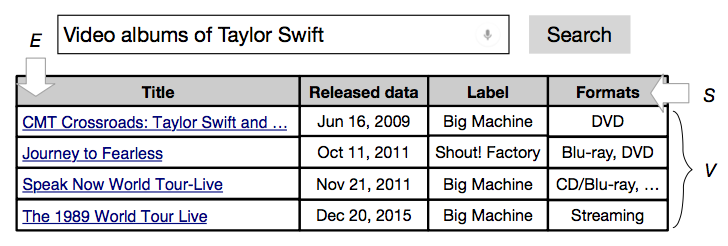} 
   \vspace*{-1.25\baselineskip}
   \caption{Answering a search query with an on-the-fly generated table, consisting of core column entities $E$, table schema $S$, and data cells $V$.}
   \vspace*{-1.25\baselineskip}
\label{fig:task_2}
\end{figure}

Tables are popular on the Web because of their convenience for organizing and managing data. 
Tables can also be useful for presenting search results~\cite{Pimplikar:2012:ATQ, Yang:2014:FPK}.
Users often search for a set of things, like music albums by a singer, films by an actor, restaurants nearby, etc.  In a typical information retrieval system, the matched entities are presented as a list.  
Search, however, is often part of a larger work task, where the user might be interested in specific attributes of these entities. 
Organizing results, that is, entities and their attributes, in a tabular format facilitates a better overview.
E.g., for the query ``video albums of Taylor Swift,'' we can list the albums in a table, as shown in Fig.~\ref{fig:task_2}.

There exist two main families of methods that can return a table as answer to a keyword query by: 
(i) performing table search to find existing tables on the Web~\cite{Cafarella:2008:WEP,Cafarella:2009:DIR,Venetis:2011:RST,Pimplikar:2012:ATQ,Zhang:2018:AHT,Nguyen:2015:RSS}, or
(ii) assembling a table in a row-by-row fashion~\cite{Yang:2014:FPK} or by joining columns from multiple tables~\cite{Pimplikar:2012:ATQ}.
However, these methods are limited to returning tables that already exist in their entirety or at least partially (as complete rows/columns).
Another line of work aims to translate a keyword or natural language query to a structured query language (e.g., SPARQL), which can be executed over a knowledge base~\citep{Yahya:2012:NLQ}.  While in principle these techniques could return a list of tuples as the answer, in practice, they are targeted for factoid questions or at most a single attribute per answer entity.  More importantly, they require data to be available in a clean, structured form in a consolidated knowledge base.
Instead, we propose to generate tables on the fly in a cell-by-cell basis, by combining information from existing tables as well as from a knowledge base, such that each cell's value can originate from a different source.

In this study, we focus on \emph{relational tables} (also referred to as \emph{genuine tables}~\citep{Wang:2002:MLB, Wang:2002:DTH}), which describe a set of entities along with their attributes~\citep{Lehmberg:2016:LPC}.  
A relational table consists of three main elements: (i) the \emph{core column entities} $E$, (ii) the \emph{table schema} $S$, which consists of the table's heading column labels, corresponding to entity attributes, and (iii) \emph{data cells}, $V$, containing attribute values for each entity.  
The task of \emph{on-the-fly table generation} is defined as follows: answering a free text query with an output table, where the core column lists all relevant entities and columns correspond the attributes of those entities.
This task can naturally be decomposed into three main components: 
\begin{enumerate}
	\item \emph{Core column entity ranking}, which is about identifying the entities to be included in the core column of the table. 
	\item \emph{Schema determination}, which is concerned with finding out what should be the column headings of the table, such that these attributes can effectively summarize answer entities. 
	\item \emph{Value lookup}, which is to find the values of corresponding attributes in existing tables or in a knowledge base.
\end{enumerate}
The first subtask is strongly related to the problem of entity retrieval~\citep{Hasibi:2017:DVT}, while the second subtask is related to the problem of attribute retrieval~\citep{Kopliku:2011:TFA}.
These two subtasks, however, are not independent of each other.  We postulate that core column entity ranking can be improved by knowing the schema of the table, and vice versa, having knowledge of the core column entities can be leveraged in schema determination.
Therefore, we develop a framework in which these two subtasks can be performed iteratively and can reinforce each other. 
As for the third subtask, value lookup, the challenge there is to find a distinct value for an entity-attribute pair, with a traceable source, from multiple sources.

In summary, the main contributions of this work are as follows:
\begin{itemize}
	\item We introduce the task of on-the-fly table generation and propose an iterative table generation algorithm (Sect.~\ref{sec:ps}).
	\item We develop feature-based approaches for core column entity ranking (Sect.~\ref{sec:corecol}) and schema determination (Sect.~\ref{sec:schema}), and design an entity-oriented fact catalog for fast and effective value lookup (Sect.~\ref{sec:vl}).
	\item We perform extensive evaluation on the component level (Sect.~\ref{sec:eval}) and provide further insights and analysis (Sect.~\ref{sec:anal}).
\end{itemize}
The resources developed within this study are made publicly available at \url{https://github.com/iai-group/sigir2018-table}.

\vspace*{-0.5\baselineskip}
\section{Overview} 
\label{sec:ps}

\begin{figure}[t]
   \centering
   \includegraphics[width=0.45\textwidth]{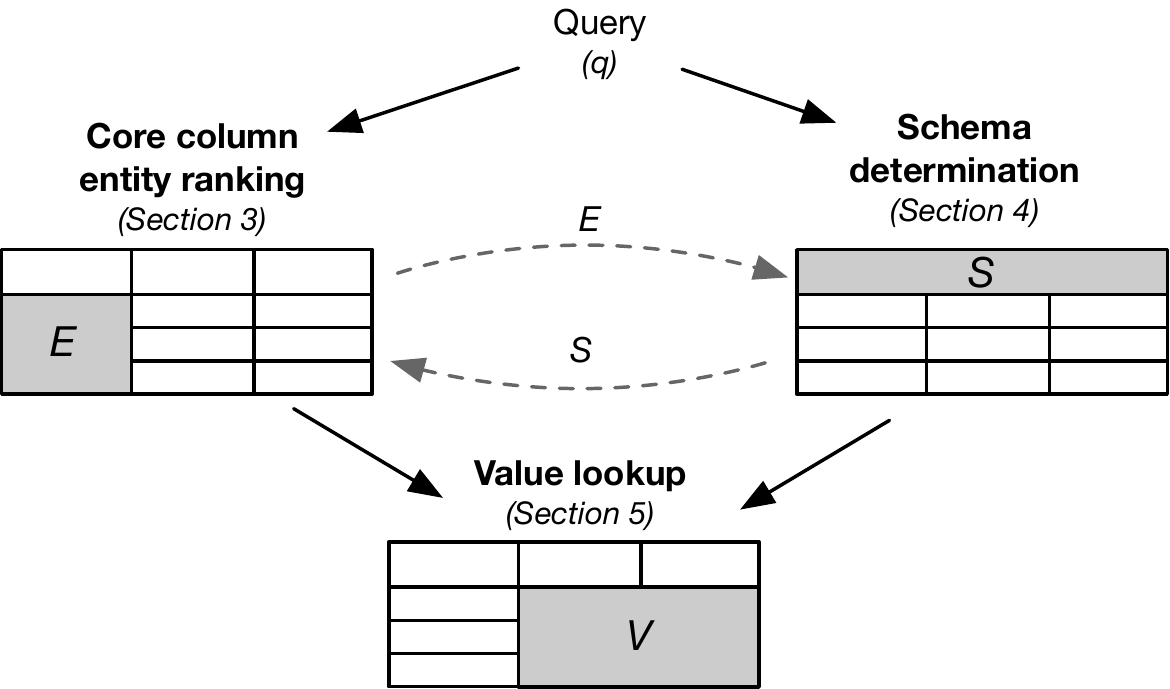} 
   \captionshrink
   \caption{Overview of our table generation approach.}
   \vspace*{-1\baselineskip}
\label{fig:overview}
\end{figure}

The objective of on-the-fly table generation is to assemble and return a relational table as the answer in response to a free text query.   
Formally, given a keyword query $q$, the task is to return a table $T=(E,S,V)$, where $E=\langle e_1,\dots e_n \rangle$ is a ranked list of core column entities, $S=\langle s_1,\dots s_m \rangle$ is a ranked list of heading column labels, and $V$ is an $n$-by-$m$ matrix, such that $v_{ij}$ refers to the value in row $i$ and column $j$ of the matrix ($i \in [1..n]$, $j \in [1..m]$).  
According to the needed table elements, the task boils down to (i) searching core column entities, (ii) determining the table schema, and (iii) looking up values for the data cells. 
Figure~\ref{fig:overview} shows how these three components are connected to each other in our proposed approach. 

%

%

%
%
\vspace*{-0.25\baselineskip}
\subsection{Iterative Table Generation Algorithm}

There are some clear sequential dependencies between the three main components: \emph{core column entity ranking} and \emph{schema determination} need to be performed before \emph{value lookup}.  Other than that, the former two may be conducted independently of and parallel to each other.  
However, we postulate that better overall performance may be achieved if core column entity ranking and schema determination would supplement each other. That is, each would make use of not only the input query, but the other's output as well.
To this end, we propose an iterative algorithm that gradually updates core column entity ranking and schema determination results. 

The pseudocode of our approach is provided in Algorithm~\ref{alg:fw}, where $\mathit{rankEntites}()$, $\mathit{rankLabels}()$, and $\mathit{lookupValues}()$ refer to the subtasks of core column entity ranking, schema determination, and value lookup, respectively.
Initially, we issue the query $q$ to search entities and schema labels, by $\mathit{rankEntites}(q,\{\})$ and $\mathit{rankLabels}(q,\{\})$.  Then, in a series of successive iterations, indexed by $t$, core column entity ranking will consider the top-$k$ ranked schema labels from iteration $t-1$ ($\mathit{rankEntites}(q,S^{t-1})$). Analogously, schema determination will take the top-$k$ ranked core column entities from the previous iteration ($\mathit{rankLabels}(q,E^{t-1})$).
These steps are repeated until some termination condition is met, e.g., the rankings do not change beyond a certain extent anymore.  We leave the determination of a suitable termination condition to future work and will use a fixed number of iterations in our experiments.
In the final step of our algorithm, we look up values $V$ using the core column entities and schema ($\mathit{lookupValues}(E^t, S^t)$).  Then, the resulting table $(E^t, S^t, V)$ is returned as output.

\begin{algorithm}[t]
\SetAlgoLined
\SetKwFunction{Range}{range}
\KwData{$q$, a keyword query}
\KwResult{$T=(E,S,V)$, a result table}
\Begin{
$E^0 \leftarrow \mathit{rankEntites}(q,\{\})$\;
$S^0 \leftarrow \mathit{rankLabels}(q,\{\})$\;
$t \leftarrow 0$ \;
 \While{$\neg terminate$}{ 
	$t \leftarrow t+1$ \;
	$E^{t} \leftarrow \mathit{rankEntites}(q,S^{t-1})$\;
	$S^{t} \leftarrow \mathit{rankLabels}(q,E^{t-1})$\;
 }
$V \leftarrow \mathit{lookupValues}(E^t, S^t)$\;
\Return{$(E^t, S^t, V)$}
}
\caption{Iterative Table Generation}
\vspace*{-0.25\baselineskip}
\label{alg:fw}
\end{algorithm}


\vspace*{-0.25\baselineskip}
\subsection{Data Sources}
\label{sub:td}

Another innovative element of our approach is that we do not rely on a single data source.  We combine information both from a collection of existing tables, referred to as the \emph{table corpus}, and from a knowledge base.
We shall assume that there is some process in place that can identify relational tables in the table corpus, based on the presence of a core column.  We further assume that entities in the core column are linked to the corresponding entries in the knowledge base.  The technical details are described in Sect.~\ref{sec:expsetup}.
Based on the information stored about each entity in the knowledge base, we consider multiple entity representations: (i) \emph{all} refers to the concatenation of all textual material that is available about the entity (referred to as ``catchall'' in~\citep{Hasibi:2017:DVT}), (ii) \emph{description} is based on the entity's short textual description (i.e., abstract or summary), and (iii) \emph{properties} consists of a restricted set of facts (property-value pairs) about the entity.
We will use DBpedia in our experiments, but it can be assumed, without loss of generality, that the above information is available in any general-purpose knowledge base.

\section{Core column entity ranking}
\label{sec:corecol}

In this section, we address the subtask of \emph{core column entity ranking}: given a query, identify entities that should be placed in the core column of the generated output table.
This task is closely related to the problem of ad hoc entity retrieval.    
Indeed, our initial scoring function is based on existing entity retrieval approaches.  However, this scoring can be iteratively improved by leveraging the identified table schema.
Our iterative scoring function combines multiple features as ranking signals in a linear fashion:
\begin{equation}
	\mathit{score}_t(e,q) = \sum_i w_{i}\phi_{i}(e,q,S^{t-1}) ~,
	\label{eq:ccer}
\end{equation}
where $\phi_i$ is a ranking feature and $w_i$ is the corresponding weight.
In the first round of the iteration ($t=0$), the table schema is not yet available, thus $S^{t-1}$ by definition is an empty list.  For later iterations ($t>0$), $S^{t-1}$ is computed using the methods described in Sect.~\ref{sec:schema}.  
For notational convenience, we shall write $S$ to denote the set of top-$k$ schema labels from $S^{t-1}$. 
In the remainder of this section, we present the features we developed for core column entity ranking; see Table~\ref{tbl:features_entity} for a summary.

\vspace*{-0.25\baselineskip}
\subsection{Query-based Entity Ranking}
\label{sec:corecol:adhoc}

Initially, we only have the query $q$ as input.  We consider term-based and semantic matching as features.

\vspace*{-0.25\baselineskip}
\subsubsection{Term-based matching}

There is a wide variety of retrieval models for term-based entity ranking~\citep{Hasibi:2017:DVT}. 
We rank document-based entity representations using Language Modeling techniques.  Despite its simplicity, this model has shown to deliver competitive performance~\citep{Hasibi:2017:DVT}.  Specifically, following~\citep{Hasibi:2017:DVT}, we use the \emph{all} entity representation, concatenating all textual material available about a given entity. 

\begin{figure}[t]
   \centering
   \includegraphics[width=0.3\textwidth]{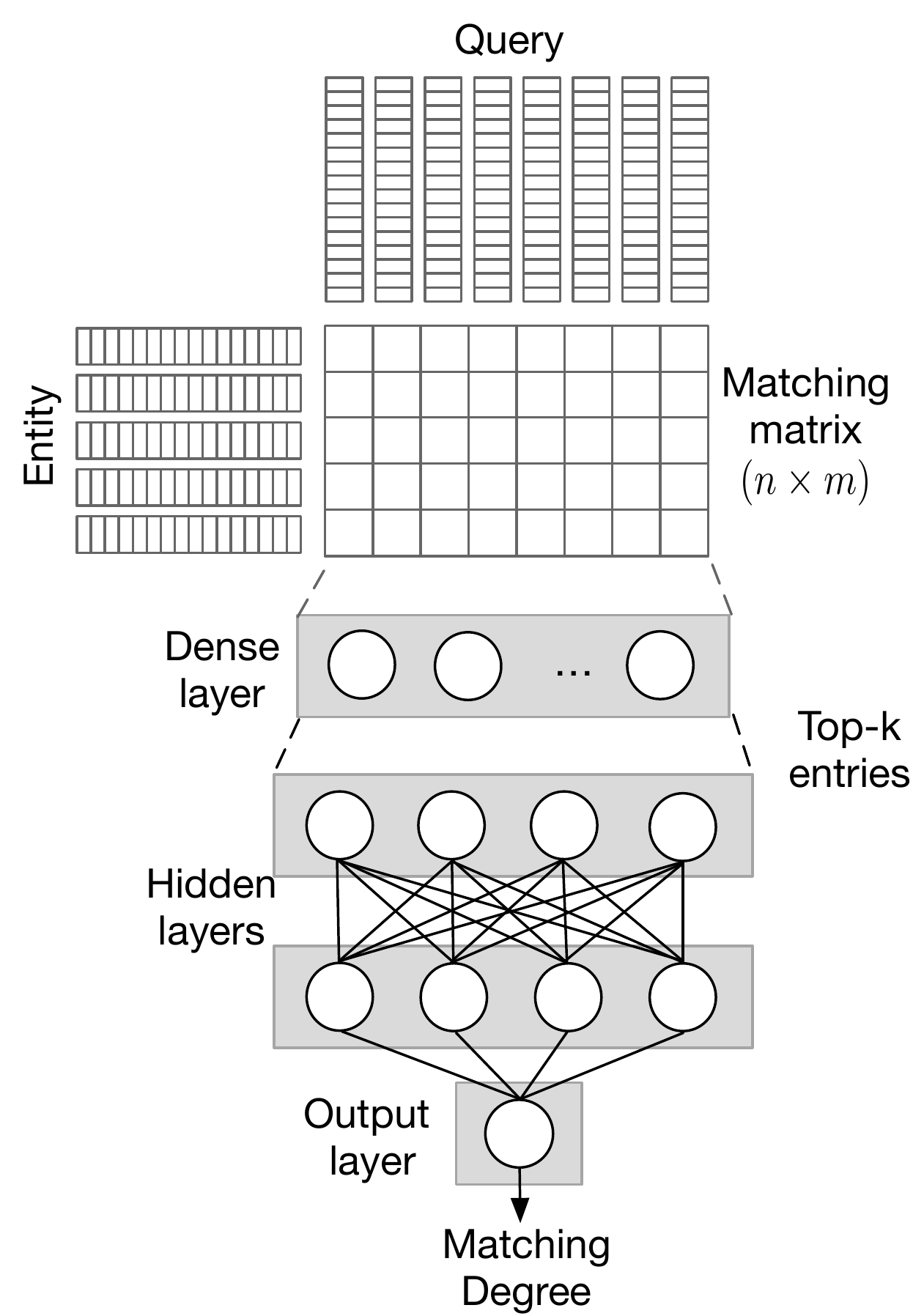} 
   \captionshrink
   \caption{Architecture of the DRRM\_TKS deep semantic matching method.}
   \vspace*{-1\baselineskip}
\label{fig:deep}
\end{figure}

\vspace*{-0.25\baselineskip}
\subsubsection{Deep semantic matching}
\label{subsubsec:dsmf}

We employ a deep semantic matching method, referred to as DRRM\_TKS~\citep{Fan:2017:MTD}. It is an enhancement of DRRM~\citep{Guo:2016:DRM} for short text, where the matching histograms are replaced with the top-$k$ strongest signals. Specifically, the entity and the query are represented as sequences of embedding vectors, denoted as $e=[w_1^e, w_2^e, ..., w_n^e]$ and $q=[w_1^q, w_2^q, ..., w_m^q]$.  An $n \times m$ matching matrix $M$ is computed for their joint representation, by setting $M_{ij}= w_i^e \cdot (w_j^q)^\intercal$. 
The values of this matrix are used as input to the dense layer of the network.  
Then, the top-$k$ strongest signals, based on a softmax operation, are selected and fed into the hidden layers.  The output layer computes the final matching score between the query and entity.
The architecture of DRRM\_TKS is shown in Fig.~\ref{fig:deep}.

We instantiate this neural network with two different entity representations: (i) using the entity's textual description, $e_d$, and (ii) using the properties of the entity in the knowledge base, $e_p$.
The matching degree scores computed using these two representations, $DRRM\_TKS(q, e_d)$ and $DRRM\_TKS(q, e_p)$, are used as ranking features $\phi_2$ and $\phi_3$, respectively.

\vspace*{-0.25\baselineskip}
\subsection{Schema-assisted Entity Ranking}
\label{sec:corecol:schema}

After the first iteration, core column entity ranking can be assisted by utilizing the determined table schema from the previous iteration.  
We present a number of additional features that incorporate schema information.

\vspace*{-0.25\baselineskip}
\subsubsection{Deep semantic matching}
We employ the same neural network as before, in Sect.~\ref{subsubsec:dsmf}, to compute semantic similarity by considering the table schema.  Specifically, all schema labels in $S$ are concatenated into a single string $s$. 
For the candidate entities, we keep the same representations as in Sect.~\ref{subsubsec:dsmf}.  By comparing all  schema labels $s$ against the entity, we obtain the schema-assisted deep features $DRRM\_TKS(s, e_d)$ and $DRRM\_TKS(s, e_p)$. 
Additionally, we combine the input query with the schema labels, $q \oplus s$, and match it against a combined representation of the entity, $e_d \oplus e_p$, where $\oplus$ refers to the string concatenation operation.  The resulting matching score is denoted as $DRRM\_TKS(q \oplus s, e_d \oplus e_p)$.

%
%

%
\begin{table}[t]
  \centering
  \caption{Features used for core column entity retrieval.}
  \captionshrink
  \begin{tabular}{llll}
    \toprule
    \multicolumn{3}{l}{\textbf{Feature}} & \textbf{Iter. ($t$)} \\
    \midrule
    \multicolumn{4}{l}{\emph{Term-based matching}} \\  
    & $\phi_1$: & $LM(q, e_a)$ & $\geq 0$ \\
    \multicolumn{4}{l}{\emph{Deep semantic matching}} \\  
    & $\phi_2$: & $DRRM\_TKS(q, e_d)$ & $\geq 0$ \\
    & $\phi_3$: & $DRRM\_TKS(q, e_p)$ & $\geq 0$ \\
    & $\phi_4$: & $DRRM\_TKS(s, e_d)$ & $\geq 1$ \\
    & $\phi_5$: & $DRRM\_TKS(s, e_p)$ & $\geq 1$ \\
    & $\phi_6$: & $DRRM\_TKS(q \oplus s, e_d \oplus e_p)$ & $\geq 1$ \\
    \multicolumn{4}{l}{\emph{Entity-schema compatibility}} \\      
    & $\phi_7$: & $ESC(S,e)$ & $\geq 1$ \\    
    \bottomrule
  \end{tabular}
  \label{tbl:features_entity}
  \captionshrink
\end{table}
%
%


\vspace*{-0.25\baselineskip}
\subsubsection{Entity-schema compatibility}
\label{subsub:scf}

Intuitively, core column entities in a given table are from the same semantic class, for example, athletes, digital products, films, etc.  We aim to capture their semantic compatibility with the table schema, by introducing a measure called \emph{entity-schema compatibility}.


We compare the property labels of core column entities $E$ against schema $S$ to build the compatibility matrix $C$.  Element $C_{ij}$ of the matrix is a binary indicator between the $j$th schema label and the $i$th entity, which equals to 1 if entity $e_i$ has property $s_j$.
To check if an entity has a given property, we look for evidence both in the knowledge base and in the table corpus. 
Formally: 
\begin{equation*}
    C_{ij} = 
	\begin{cases}
    	1, & \text{if~}\mathit{match}_{KB}(e_i,s_j) \vee \mathit{match}_{TC}(e_i,s_j) \\ 
    	0,              & \text{otherwise} ~.
	\end{cases}
\end{equation*}
where $\mathit{match}_{KB}(e_i,s_j)$ and $\mathit{match}_{TC}(e_i,s_j)$ are binary indicator functions. The former is true if entity $e_i$ has property $s_j$ in the knowledge base, the latter is true if there exists a table in the table corpus where $e_i$ is a core column entity and $s_j$ is a schema label.
Then, the entity-schema compatibility score, which is used as ranking feature $\phi_7$, is computed as follows:
\begin{equation*}
	ESC(S,e_i) = \frac{1}{|S|}\sum_j C_{ij} ~.
\end{equation*}
For example, for query ``Apollo astronauts walked on the Moon'' and schema $\{$country, date of birth, time in space, age at first step, ...$\}$, the ESC scores of entities Alan Shepard, Charles Duke, and Bill Kaysing are 1, 0.85, and 0.4, respectively. The former two are Apollo astronauts who walked on the Moon, while the latter is a writer claiming that the six Apollo Moon landings were a hoax.

\vspace*{-0.25\baselineskip}
\section{Schema determination}
\label{sec:schema}

In this section, we address the subtask of \emph{schema determination}, which is to return a ranked list of labels to be used as heading column labels (\emph{labels}, for short) of the generated output table.  The initial ranking is based on the input query only.  Then, this ranking is iteratively improved by also considering the core column entities.  
Our scoring function is defined as follows:
\begin{equation}
	\mathit{score}_t(s,q) = \sum_i w_{i}\phi_{i}(s,q,E^{t-1}),
	\label{eq:sd}
\end{equation}
where $\phi_i$ is a ranking feature with a corresponding weight $w_i$.
For the initial ranking ($t=0$), core column entities are not yet available, thus $E^{t-1}$ is an empty list.  
For successive iterations ($t>0$), $E^{t-1}$ is computed using the methods described in Sect.~\ref{sec:corecol}.
Since we are only interested in the top-$k$ entities, and not their actual retrieval scores, we shall write $E$ to denote the set of top-$k$ entities in $E^{t-1}$.
Below, we discuss various feature functions $\phi_{i}$ for this task, which are also summarized in Table~\ref{tbl:features_schema}.

\begin{table}[t]
  \centering
  \caption{Features used for schema determination.}
  \captionshrink
  \begin{tabular}{llll}
    \toprule
    \multicolumn{3}{l}{\textbf{Feature}} & \textbf{Iter. ($t$)} \\
    \midrule
    \emph{Column population}  
    & $\phi_1$: & $P(s|q)$ & $\geq 0$ \\
    & $\phi_2$: & $P(s|q,E)$ & $\geq 1$ \\
    \emph{Deep semantic matching}  
    & $\phi_3$: & $DRRM\_TKS(s,q)$ & $\geq 0$ \\
    \emph{Attribute retrieval}
    & $\phi_4$: & $AR(s,E)$ & $\geq 1$ \\    
    \emph{Entity-schema compatibility}      
    & $\phi_5$: & $ESC(s,E)$ & $\geq 1$ \\    
    \bottomrule
  \end{tabular}
  \label{tbl:features_schema}
\end{table}

\vspace*{-0.25\baselineskip}
\subsection{Query-based Schema Determination}

At the start, only the input query $q$ is available for ranking labels. 
%
%
To collect candidate labels, we first search for tables in our table corpus that are relevant to the query.  We let $\mathcal{T}$ denote the set of top-$k$ ranked tables.  Following~\citep{Zhang:2017:ESA}, we use BM25 to rank tables based on their textual content.
Then, the column heading labels are extracted from these tables as candidates: $\mathcal{S} = \{s | s \in T_S, T \in \mathcal{T} \}$.

\vspace*{-0.25\baselineskip}
\subsubsection{Column population}
\label{subsub:cpf}

\citet{Zhang:2017:ESA} introduce the task of \emph{column population}: generating a ranked list of column labels to be added to the column headings of a given seed table.  We can adapt their method by treating the query as if it was the caption of the seed table.  Then, the scoring of schema labels is performed according to the following probabilistic formula:
\begin{equation*}
	P(s|q) = \sum_{T \in \mathcal{T}} P(s|T)P(T|q) ~,
\label{eq:ptecl}
\end{equation*}
where related tables serve as a bridge to connect the query $q$ and label $s$.  Specifically, $P(s|T)$ is the likelihood of the schema label given table $T$ and is calculated based on the maximum edit distance~\cite{Lehmberg:2015:MSJ}, $dist$,\footnote{Note that despite the name used in~\cite{Lehmberg:2015:MSJ}, it is in fact a similarity measure.} between the $s$ and the schema labels of T:
\begin{equation}
    P(s|T) = 
	\begin{cases}
    	1, & \max_{s' \in T_S}{dist(s, s')} \ge \gamma \\
    	0,              & \text{otherwise} ~.
	\end{cases}
	\label{eq:pst}
\end{equation}
The probability $P(T|q)$ expresses the relevance of $T$ given the query, and is set proportional to the table's retrieval score (here: BM25).

%

\vspace*{-0.25\baselineskip}
\subsubsection{Deep semantic matching}
We employ the same neural network architecture as in Sect.~\ref{subsubsec:dsmf} for comparing labels against the query. 
For training the network, we use our table corpus and treat table captions as queries.  All caption-label pairs that co-occur in an existing table are treated as positive training instances.  Negative training instances are sampled from the table corpus by selecting candidate labels that do not co-occur with that caption. 
The resulting matching score, $DRRM\_TKS(s,q)$, is used as feature $\phi_3$.

\vspace*{-0.25\baselineskip}
\subsection{Entity-assisted Schema Determination}
\label{sub:eas}

After the initial round, schema determination can be assisted by considering the set of top-$k$ core column entities, $E$. 
The set of candidate labels, from before, is expanded with (i) schema labels from tables that contain any of the entities in $E$ in their core column and (ii) the properties of $E$ in the knowledge base.


\vspace*{-0.25\baselineskip}
\subsubsection{Entity enhanced column population}
We employ a variant of the column population method from~\citep{Zhang:2017:ESA} that makes use of core column entities: 
%
\begin{equation*}
	P(s|q,E) = \sum_T P(s|T) P(T|q,E) ~.
\label{eq:ptecl}
\end{equation*}
The schema label likelihood $P(s|T)$ is computed the same as before, cf. Eq.~\eqref{eq:pst}.
The main difference is in the table relevance estimation component, which now also considers the core column entities:
\begin{equation*}
	P(T|q,E) = \frac{P(T|E)P(T|q)}{P(T)^2} ~.
\end{equation*}
Here, $P(T|E)$ is the fraction of the core column entities covered by a related table, i.e., $|T_E \cap E|/|E|$, and $P(T|q)$ is the same as in \S\ref{subsub:cpf}.

\vspace*{-0.25\baselineskip}
\subsubsection{Attribute retrieval}
\label{sec:schema:ar}

Attribute retrieval refers to the task of returning a ranked list of attributes that are relevant given a set of entities~\citep{Kopliku:2011:TFA}.
Using the core column entities as input, we employ the method proposed by \citet{Kopliku:2011:TFA}, which is a linear combination of several features:
\begin{equation*}
	AR(s,E) = 
	\frac{1}{|E|}\sum_{e \in E}\big( match(s,e,T)+ drel(d,e)+ sh(s,e) + kb(s,e) \big ) ~.
\end{equation*}
The components of this formula are as follows:
\begin{itemize}
	\item $match(s, e, T)$ compares the similarity between an entity and a schema label with respect to a given table $T$. We take $T$ to be the table that is the most relevant to the query ($\arg\max_{T\in \mathcal{T}} P(T|q)$).
		This matching score is the difference between the table match score and shadow match score:
		\begin{equation*}
			match(s, e, T) = match(e, T) - match(e, shadow(a)) ~.
		\end{equation*}
		The table match score is computed by representing both the entity and table cells $T_{xy}$ as term vectors, then taking the maximum cosine distance between the two:
		\begin{equation*}
			match(e, T) = max_{T_{xy} \in T}cos(e, T_{xy}) ~.
		\end{equation*}
		For latter component, the notion of a \emph{shadow area} is introduced: $shadow(a)$ is set of cells in the table that are in the same row with $e$ or are in the same column with the $s$.  Then, the shadow match score is estimated as:  
			\begin{equation*}
				match(e, shadow(a)) = \max_{T_{xy} \in \,\mathit{shadow}(a)}\cos(e, T_{xy}) ~.
			\end{equation*}
	\item $drel(d,e)$ denotes the relevance of the document $d$ that contains $T$: 
		\begin{equation*}
			drel(e) = \frac{\#results-rank(d)}{\#results} ~,	
		\end{equation*}
		where $\#results$ is the number of retrieved results for entity $e$ and $rank(d)$ is the rank of document $d$ within this list. 
	\item $sh(s,e)$ corresponds to the number of search results returned by a Web search engine to a query ``$\langle s \rangle$ of $\langle e \rangle$,'' where $s$ and $e$ are substituted with the label and entity, respectively.  If the base-10 logarithm of the number of hits exceeds a certain threshold ($10^6$ in~\citep{Kopliku:2011:TFA}) then the feature takes a value of $1$, otherwise it is $0$.
	\item $kb(s,e)$ is a binary score indicating whether label $s$ is a property of entity $e$ in the knowledge base (i.e., $s \in e_p$).

\end{itemize}


\vspace*{-0.25\baselineskip}
\subsubsection{Entity-schema compatibility}

Similar to Sect.~\ref{subsub:scf}, we employ the entity-schema compatibility feature for schema determination as well. As before, $C$ is a compatibility matrix, where $C_{ij}$ denotes whether entity $e_i$ has property $s_j$.  The ESC score is then computed as follows:
\begin{equation*}
	ESC(s_j, E) = \frac{1}{|E|}\sum_i C_{ij} ~.
\end{equation*}
%
\vspace*{-1\baselineskip}
\section{Value Lookup}
\label{sec:vl}

Having the core column entities and the schema determined, the last component in our table generation approach is concerned with the retrieval of the data cells' values.  Formally, for each row (entity) $i \in [1..n]$ and column (schema label) $j \in [1..m]$, our task is to find the value $V_{ij}$.  
This value may originate from an existing table in our table corpus or from the knowledge base.  The challenges here are twofold: (i) how to match the schema label $s_j$ against the labels of existing tables and knowledge base predicates, and (ii) how to deal with the case when multiple, possibly conflicting values may be found for a given cell.

We go about this task by first creating a catalogue $\mathcal{V}$ of all possible cell values.  Each possible cell value is represented as a quadruple $\langle e, s, v, p \rangle$, where $e$ is an entity, $s$ is a schema label, $v$ is a value, and $p$ is provenance, indicating the source of the information.  The source may be a knowledge base fact or a particular table in the table corpus.  
An entity-oriented view of this catalog is a filtered set of triples where the given entity stands as the first component of the quadruple:
$e_V = \{ \langle s, v, p \rangle | \langle e, s, v, p \rangle \in \mathcal{V} \}$.
We select a single value for a given entity $e$ and schema label $s$ according to:
\begin{equation*}
	\mathit{score}(v,e,s,q) = \max_{\substack{\langle s', v, p \rangle \in e_V\\ \mathit{match}(s,s')}} \mathit{conf}(p,q) ~,
\end{equation*}
%
%
where $\mathit{match}(s,s')$ is a soft string matching function (detailed in Sect.~\ref{sub:sn}) and $\mathit{conf}(p,q)$ is the confidence associated with provenance $p$.  
Motivated by the fact that the knowledge base is expected to contain high-quality manually curated data, we set the confidence score such that the knowledge base is always given priority over the table corpus.  If the schema label does not match any predicate from the knowledge base, then we chose the value from the table that is the most relevant to the query.  That is, $\mathit{conf}(p,q)$ is based on the corresponding table's relevance score; see Sect.~\ref{sec:eval:vl} for the details.
Notice that we pick a single source for each value rather than aggregating evidence from multiple sources. The reason for that is that on the user interface, we would like to display a single traceable source where the given value originates from.

\section{Experimental Setup}
\label{sec:expsetup}

Queries, dataset, data preprocessing methods and relevance assessments are introduced in this section.

\subsection{Test Queries}

We use two sets of queries in our experiments:

\begin{description}
	\item[QS-1] We consider list type queries from the DBpedia-Entity v2 test collection~\cite{Hasibi:2017:DVT}, that is, queries from SemSearch LS, TREC Entity, and QALD2.  Out of these, we use the queries that have at least three highly relevant entities in the ground truth. This set contains 119 queries in total.
	\item[QS-2] The RELink Query Collection~\citep{Saleiro:2017:RRF} consists of  600 complex entity-relationship queries that are answered by entity tuples.  That is, the answer table has two or three columns (including the core entity column) and all cell values are entities.  The queries and corresponding relevance judgments in this collection are obtained from Wikipedia lists that contain relational tables.  For each answer table, human annotators were asked to formulate the corresponding information need as a natural language query, e.g., ``find peaks above 6000m in the mountains of Peru.''
\end{description}
For both sets, we remove stop words and perform spell correction.

\subsection{Data Sources}
\label{sec:data}

We rely on two main data sources simultaneously: a knowledge base and a table corpus.

\subsubsection{Knowledge base} 
The knowledge base we use is DBpedia (version 2015-10). We consider entities for which a short textual description is given in the \texttt{dbo:abstract} property (4.6M in total).  We limit ourselves to properties that are extracted from Wikipedia infoboxes.


\subsubsection{Table corpus}

We use the WikiTables corpus~\cite{Bhagavatula:2015:TEL}, which contains 1.65M tables extracted from Wikipedia. The mean number of rows is 11 and the median is 5. For columns, the mean is 5 and the median is 4. We preprocess  tables as follows. 
For each cell that contains a hyperlink we check if it points to an entity that is present in DBpedia.  If yes, we use the DBpedia identifier of the linked entity as the cell's content (with redirects resolved); otherwise, we replace the link with the anchor text (i.e., treat it as a string). 

Further, each table is classified as relational or non-relational according to the existence of a core entity column and the size of the table. 
We set the following conditions for detecting the core column of a table:
(i) the core column should contain the most entities compared to other columns;
(ii) if there are more than one columns that have the highest number of entities, then the one with lowest index, i.e., the leftmost one, is regarded as the core column;
(iii) the core column must contain at least two entities.
Tables without a core column or having less than two rows or columns are regarded as non-relational. 
In the end, we classify the WikiTables corpus into 973,840 relational and 678,931 non-relational tables.   Based on a random sample of 100 tables from each category, we find that all the sampled tables are correctly classified. 

%
%
%
%
%
%
%
%
%


\vspace*{-0.5\baselineskip}
\subsection{Schema Normalization}
\label{sub:sn}
Different schema labels may be used for expressing the same meaning, e.g., ``birthday'' vs. ``day of birth'' or ``nation'' vs. ``country.'' For the former case, where similar terms are used, we employ a FastJoin match~\cite{Wang:2014:ESS} to normalize the strings (with stopwords removed). Specifically, we take the maximum edit distance as in~\cite{Lehmberg:2015:MSJ} to measure string similarity. When it exceeds a threshold of $\delta$, we regard them as the same label. We set $\delta$ as 0.8 which is consistent with~\cite{Lehmberg:2015:MSJ}, where headings are matched for table column join. For the latter case, where different terms are used, we consider predicates connecting the same subject and object as  synonyms.  These pairs are then checked and erroneous ones are eliminated manually. 
Whenever schema labels are compared in the paper, we use their normalized versions.



%

%

\vspace*{-0.5\baselineskip}
\subsection{Relevance Assessments}
\label{sub:ra}


For QS-1, we consider the highly relevant entities as the ground truth for the core column entity ranking task. For the task of schema determination, we annotated all candidate labels using crowdsourcing.  
Specifically, we used the CrowdFlower platform and presented annotators with the query, three example core column entities, and a label, and asked them to judge the relevance of that label on a three point scale: highly relevant, relevant, or non-relevant. Each query-entity-label triple was annotated by at least three and at most five annotators. The labelling instructions were as follows: a label is highly relevant if it corresponds to an essential table column for the given query and core column entities; a label is relevant when it corresponds to a property shared by most core column entities and provides useful information, but it is not essential for the given query; a label is non-relevant otherwise (e.g., hard to understand, not informative, not relevant, etc.). We take the majority vote to decide the relevance of a label. Statistically, we have 7000 triples annotated, and on average, there are 4.2 highly relevant labels, 1.9 relevant labels, and 49.4 non-relevant labels for each query.  The Fleiss' Kappa test statistics for inter-annotator agreement is 0.61, which is considered as substantial agreement~\cite{Fleiss:1971:MNS}. 
For the value lookup task, we sampled 25 queries and fetched values from the table corpus and the knowledge base. We again set up a crowdsourcing experiment on CrowdFlower for annotation. Given a query, an entity, a schema label, a value, and a source (Wikipedia or DBpedia page), three to five annotators were asked to validate if the value can be found and whether it is correct, according to the provided source.  Overall, 14,219 table cell values were validated. The total expense of the crowdsourcing experiments was \$560.
	
\emph{QS-2}: Since for this query set we are given the ground truth in a tabular format, based on existing Wikipedia tables, we do not need to perform additional manual annotation.  The main entities are taken as the ground truth for the core column entity ranking task, heading labels are taken as the ground truth for the schema determination task, and the table cells (for a sample of 25 queries) are taken as the ground truth for the value lookup task.
%
%
%

\vspace*{-0.5\baselineskip}
\subsection{Evaluation Measures}
We evaluate core column entity ranking and schema determination in terms of Normalized Discounted Cumulative Gain (NDCG) at cut-off points 5 and 10.  The value lookup task is measured by Mean Average Precision (MAP) and Mean Reciprocal Rank (MRR). 
To test significance, we use a two-tailed paired t-test and write $\dag$/$\ddag$ to denote significance at the 0.05 and 0.005 levels, respectively.
 
\section{Experimental Evaluation}
\label{sec:eval}

We evaluate the three main components of our approach, core column entity ranking, schema determination, and value lookup, and assess the effectiveness of our iterative table generation algorithm. 



\subsection{Core Column Entity Ranking}

We discuss core column entity ranking results in two parts: (i) using only the query as input and (ii) leveraging the table schema as well.

\subsubsection{Query-based Entity Ranking}
The results are reported in top block of Table~\ref{tbl:er_qs}.
The following methods are compared:

\begin{description}
	\item[LM] For term-based matching we use Language Modeling with Dirichlet smoothing, with the smoothing parameter set to 2000, following~\citep{Hasibi:2017:DVT}. This method is also used for obtaining the candidate set (top 100 entities per query) that are re-ranked by the methods below. 
	\item[DRRM\_TKS] We train the deep matching model using 5-fold cross-validation. We use a four-layer architecture, with 50 nodes in the input layer, two hidden layers in the feed forward matching networks, and one output layer. The optimizer is ADAM~\citep{Kingma:2014:AMS}, with hinge loss as the loss function. We set the learning rate to 0.0001 and we report the results after 50 iterations.\footnote{We also experimented with C-DSSM and DSSM. However, their overall performance was much lower than that of DRRM\_TKS for this task.} We employ two instantiations of this network, using entity descriptions ($e_d$) and entity properties ($e_p$) as input.
	\item[Combined] We combine the previous three methods, with equal weights, using a linear combination (cf. Eq.~\ref{eq:ccer}).  Later, in our analysis in Sect.~\ref{sec:anal:param}, we will also experiment with learning the weights for the combination.
\end{description}
On the first query set, QS-1, LM performs best of the single rankers. Combining it with deep features results in 16\% and 9\% relative improvement for NDCG@5 and NDCG@10, respectively. 
On QS-2, a slightly different picture emerges.  The best individual ranker is DRRM\_TKS using entity properties.  Nevertheless, the Combined method still improves significantly over the LM baseline.

\begin{table}[t]
\center
\caption{Core column entity ranking results. The top block of the table uses only the keyword query as input. The bottom block of the table uses the table schema; Round \#1--\#3 rely on automatically determined schema, while the Oracle method uses the ground truth schema.  Statistical significance for query-based entity ranking is compared against LM, for schema-assisted entity ranking is compared against the Combined method.}
  \captionshrink
	\begin{tabular}{lc@{~}c@{~~}cc@{~~}c}
	\toprule
    	& \multicolumn{2}{c}{\textbf{QS-1}} && \multicolumn{2}{c}{\textbf{QS-2}} \\
    \cline{2-3} \cline{5-6} 
    \textbf{Method} & \textbf{\footnotesize NDCG@5} & \textbf{\footnotesize NDCG@10} & &
    \textbf{\footnotesize NDCG@5} & \textbf{\footnotesize NDCG@10}  \\
	\midrule
	\multicolumn{6}{l}{\emph{Query-based Entity Ranking (Round \#0)}} \\
	\midrule
	LM & 0.2419 & 0.2591 && 0.0708 & 0.0823 \\
	DRRM\_TKS $(e_d)$ & 0.2015 & 0.2028 && 0.0501 & 0.0540 \\
	DRRM\_TKS $(e_p)$ & 0.1780 & 0.1808 && \textbf{0.1089$^\ddag$} & \textbf{0.1083$^\ddag$} \\
	Combined & \textbf{0.2821$^\dag$} & \textbf{0.2834} && 0.0852$^\ddag$ & 0.0920$^\dag$ \\
	\midrule
	\multicolumn{6}{l}{\emph{Schema-assisted Entity Ranking}} \\
	\midrule	
	Round \#1 & 0.3012 & 0.2892 && 0.1232$^\ddag$ & 0.1201$^\ddag$ \\ 
	Round \#2 & 0.3369$^\ddag$ & 0.3221$^\ddag$ && 0.1307$^\ddag$ & 0.1264$^\ddag$ \\ 
	Round \#3 & 0.3445$^\ddag$ & 0.3250$^\ddag$ && 0.1345$^\ddag$ & 0.1270$^\ddag$\\ 
	Oracle & \textbf{0.3518$^\ddag$} & \textbf{0.3355$^\ddag$} && \textbf{0.1587$^\ddag$} & \textbf{0.1555$^\ddag$} \\ 
	\bottomrule	
	\end{tabular}	
\label{tbl:er_qs}
\vspace*{-\baselineskip}
\end{table}

\subsubsection{Schema-assisted Entity Ranking}

Next, we also consider the table schema for core column entity ranking. The results are presented in the bottom block of Table~\ref{tbl:er_qs}.
Note that on top of to the three features we have used before, we have four additional features (cf. Table~\ref{tbl:features_entity}).  As before, we use uniform weight for all features.
We report results for three additional iterations, Rounds \#1--\#3, where the schema is taken from the previous iteration of the schema determination component.  Further, we report on an Oracle method, which uses the ground truth schema.  In all cases, we take the top 10 schema labels ($k=10$); we analyze the effect of using different $k$ values in Sect.~\ref{sub:ite}.
These methods are to be compared against the Combined method, which corresponds to Round \#0.
We find that our iterative algorithm is able to gradually improve results, in each iteration, for both of the query sets and evaluation metrics; with the exception of QS-1 in Round \#1, all improvements are highly significant.
Notice that the relative improvement made between Round \#0 and Round \#3 is substantial: 22\% and 86\% in terms of NDCG@5 for QS-1 and QS-2, respectively.

\subsection{Schema Determination}

Schema determination results are presented in two parts: (i) using only the query as input and (ii) also leveraging core column entities.

\begin{table}[t]
\center
\caption{Schema determination results. The top block of the table uses only the keyword query as input. The bottom block of the table uses the core column entities as well; Round \#1--\#3 rely on automatic entity ranking, while the Oracle method uses the ground truth entities.  Statistical significance for query-based schema determination is compared against CP, for entity-assisted entity ranking is compared against the Combined method.}
  \captionshrink
	\begin{tabular}{lccccc}
	\toprule
    	& \multicolumn{2}{c}{\textbf{QS-1}} && \multicolumn{2}{c}{\textbf{QS-2}} \\
    \cline{2-3} \cline{5-6} 
    \textbf{Method} & \textbf{\footnotesize NDCG@5} & \textbf{\footnotesize NDCG@10} & &
    \textbf{\footnotesize NDCG@5} & \textbf{\footnotesize NDCG@10}  \\
    \midrule
	\multicolumn{6}{l}{\emph{Query-based Entity Ranking (Round \#0)}} \\
    \midrule
	CP & 0.0561 & 0.0675 && 0.1770 & 0.2092 \\
	DRRM\_TKS & 0.0380 & 0.0427 && 0.0920 & 0.1415 \\
	Combined & \textbf{0.0786$^\dag$} & \textbf{0.0878$^\dag$}  && \textbf{0.2310$^\ddag$} & \textbf{0.2695$^\ddag$} \\
	\midrule
	\multicolumn{6}{l}{\emph{Entity-assisted Schema Determination}} \\
	\midrule	
	Round \#1 & 0.1676$^\ddag$ & 0.1869$^\ddag$ && 0.3342$^\ddag$ & 0.3845$^\ddag$ \\
	Round \#2 & 0.1775$^\ddag$ & 0.2046$^\ddag$ && 0.3614$^\ddag$ & 0.4143$^\ddag$ \\
	Round \#3 & 0.1910$^\ddag$ & 0.2136$^\ddag$ && 0.3683$^\ddag$ & 0.4350$^\ddag$ \\
	Oracle & \textbf{0.2002$^\ddag$} & \textbf{0.2434$^\ddag$} && \textbf{0.4239$^\ddag$} & \textbf{0.4825$^\ddag$}\\
	\bottomrule	
	\end{tabular}	
\label{tbl:sd_qs}
\end{table}

% ...

\subsubsection{Query-based Schema Determination}

In the top block of Table~\ref{tbl:sd_qs} we compare the following three methods:

\begin{description}
	\item [CP] We employ the column population method from~\citep{Zhang:2017:ESA} to determine the top 100 labels for each query.  Following~\citep{Lehmberg:2015:MSJ}, the $\gamma$ parameter for the edit distance threshold is set to 0.8.  This method is also used for obtaining the candidate label set (top 100 per query) that is re-ranked by the methods below.
	\item [DRRM\_TKS] We use the same neural network architecture as for core column entity ranking.  For training the network, we make use of Wikipedia tables.  If an entity and a schema label co-occur in an existing Wikipedia table, then we consider it as a positive pair.  Negative training instances are generated by sampling, for each entity, a set of schema labels that do not co-occur with that entity in any existing table.  In the end, we generate a total of 10.7M training examples, split evenly between the positive and negative classes. 

	\item[Combined] We combine the above two methods in a linear fashion, with equal weights (cf. Eq.~\ref{eq:sd}).  Later, in our analysis in Sect.~\ref{sec:anal:param}, we will also experiment with learning the weights for the combination.
	
\end{description}
We find that the CP performs better than DRRM\_TKS, especially on the QS-2 query set.  The Combined method substantially and significantly outperforms both of them, with a relative improvement of 40\% and 30\% over CP in terms of NDCG@5 on QS-1 and QS-2, respectively.

\subsubsection{Entity-assisted Schema Determination}

Next, we incorporate three additional features that make use of core column entities (cf. Table~\ref{tbl:features_schema}), using uniform feature weights.  For the attribute retrieval feature (\S\ref{sec:schema:ar}), we rely on the Google Custom Search API to get search hits and use the same parameter setting (feature weights) as in~\citep{Kopliku:2011:TFA}.
For all features, we use the top 10 ranked entities (and analyze different $k$ values later, in Sect.~\ref{sub:ite}).

The results are shown in the bottom block of Table~\ref{tbl:sd_qs}.
Already Round \#1 shows a significant jump in performance compared to the Combined method (corresponding to Round \#0).  Subsequent iterations results in further improvements, reaching a relative improvement of 243\% and 159\% for Round \#3 in terms of NDCG@5 for QS-1 and QS-2, respectively. 
Judging from the performance of the Oracle method, there is  further potential for improvement, especially for QS-2.

\subsection{Value Lookup}
\label{sec:eval:vl}

For value lookup evaluation we take the core column entities and schema labels from the ground truth. This is to ensure that this component is evaluated on its own merit, without being negatively influenced by errors that incur earlier in the processing pipeline.  In our evaluation, we ignore cells that have empty values according to the ground truth (approximately 12\% of the cells have empty values in the Wikitables corpus).
The overall evaluation results are reported in Table~\ref{tbl:vl}.
We rely on two sources for value lookup, the knowledge base (KB) and the table corpus (TC).
Overall, we reach excellent performance on both query sets.  On QS-1, the knowledge base is the primary source, but the table corpus also contributes new values.  On QS-2, since all values originate from existing Wikipedia tables, using the knowledge base does not bring additional benefits. This, however, is the peculiarity of that particular dataset.  Also, according to the ground truth there is a single correct value for each cell, hence the MAP and MRR scores are the same for QS-2.

%

\begin{table}[t]
\center
\caption{Value lookup results.}
  \captionshrink
	\begin{tabular}{lccccc}
	\toprule
    	& \multicolumn{2}{c}{\textbf{QS-1}} && \multicolumn{2}{c}{\textbf{QS-2}} \\
    \cline{2-3} \cline{5-6} 
    \textbf{Source} & \textbf{MAP} & \textbf{MRR} & &
    \textbf{MAP} & \textbf{MRR}  \\
    \midrule
	KB & 0.7759 & 0.7990 & & 0.0745 & 0.0745 \\
	TC & 0.1614 & 0.1746 & & 0.9564 & 0.9564 \\
	KB+TC & 0.9270 & 0.9427 & & 0.9564 & 0.9564 \\
	\bottomrule	
	\end{tabular}	
	\captionshrink
\label{tbl:vl}
\end{table}



%


\section{Analysis}
\label{sec:anal}

In this section, we conduct further analysis to provide insights on our iterative algorithm and on feature importance.

\subsection{Iterative Algorithm}
\label{sub:ite}

\begin{figure*}[t] 
  \captionshrink
  \begin{subfigure}[t]{0.24\linewidth}
    \includegraphics[width=1\linewidth]{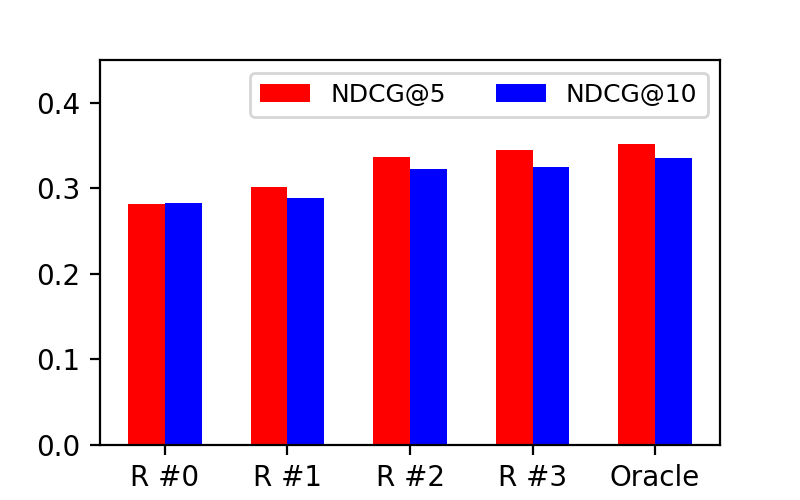} 
    \caption{CCER QS-1} 
    \label{fig:iter_a} 
  \end{subfigure}
  \begin{subfigure}[t]{0.24\linewidth}
    \includegraphics[width=1\linewidth]{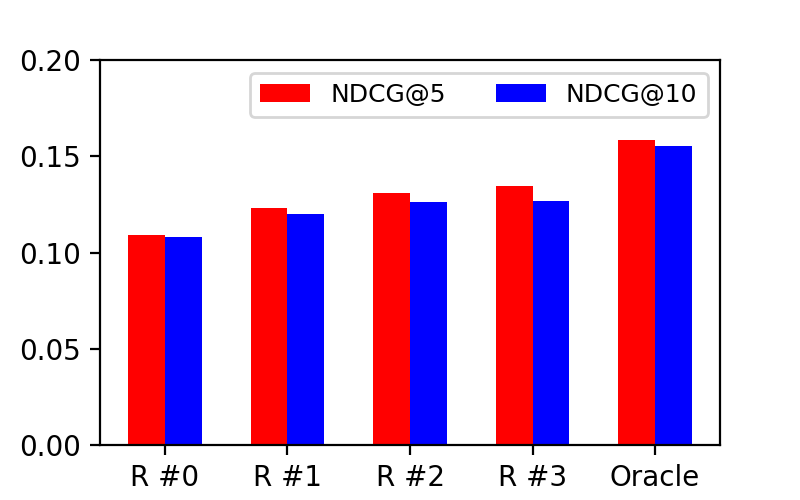} 
    \caption{CCER QS-2} 
  \end{subfigure} 
  \begin{subfigure}[t]{0.24\linewidth}
    \includegraphics[width=1\linewidth]{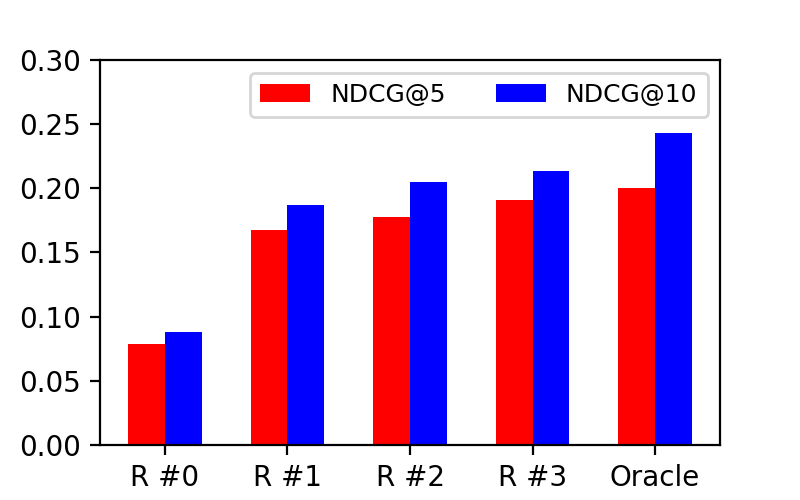} 
    \caption{SD QS-1} 
  \end{subfigure}
  \begin{subfigure}[t]{0.24\linewidth}
    \includegraphics[width=1\linewidth]{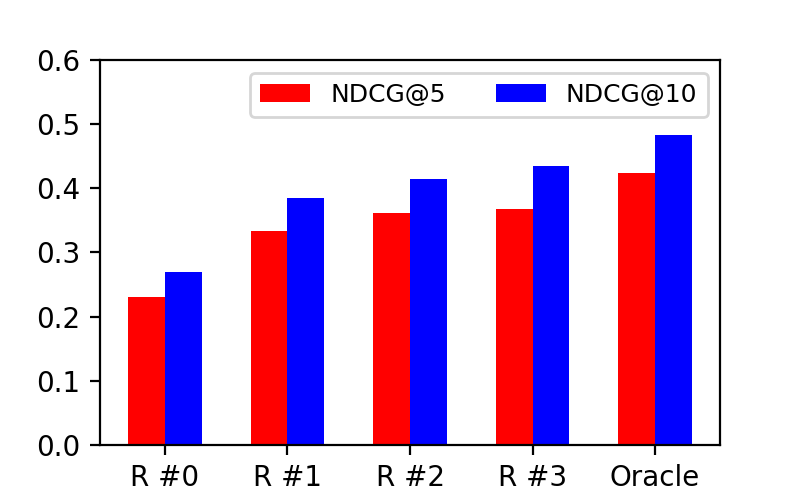} 
    \caption{SD QS-2} 
  \end{subfigure} 
  \captionshrink
  \caption{Performance change across iterations for core column entity ranking (CCER) and schema determination (SD).}
  \label{fig:iter} 
  \captionshrink
\end{figure*}

We start our discussion with Fig.~\ref{fig:iter}, which displays the overall effectiveness of our iterative algorithm on both tasks.  
Indeed, as it is clearly shown by these plots, our algorithm performs well.  The improvements are the most pronounced when going from Round \#0 to Round \#1.  Performance continues to rise with later iterations, but, as it can be expected, the level of improvement decreases over time.
The rightmost bars in the plots correspond to the Oracle method, which represents the upper limit that could be achieved, given a perfect schema determination method for core column entity ranking and vice versa.
We can observe that for core column entity ranking on QS-1 (Fig.~\ref{fig:iter_a}), has already reached this upper performance limit at iteration \#3.  For the other task/query set combinations there remains some performance to be gained.  It is left for future work to devise a mechanism for determining the number of iterations needed.  

\begin{figure*}[t] 
  \begin{subfigure}[t]{0.24\linewidth}
    \includegraphics[width=1\linewidth]{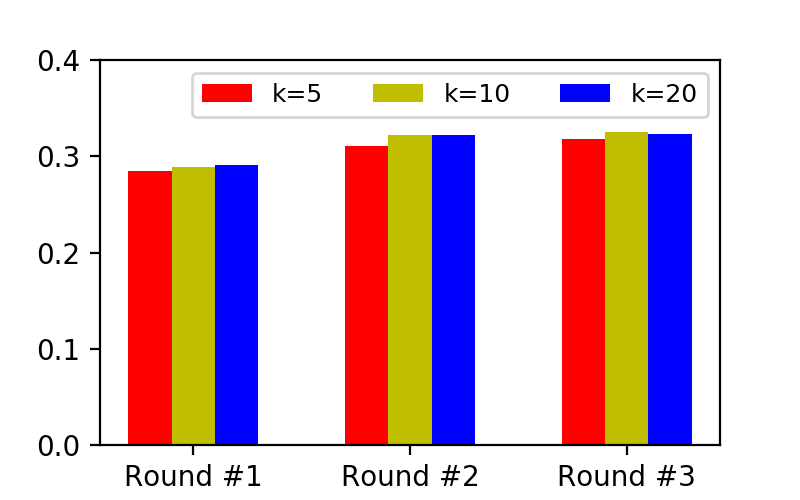} 
    \caption{CCER QS-1} 
  \end{subfigure}
  \begin{subfigure}[t]{0.24\linewidth}
    \includegraphics[width=1\linewidth]{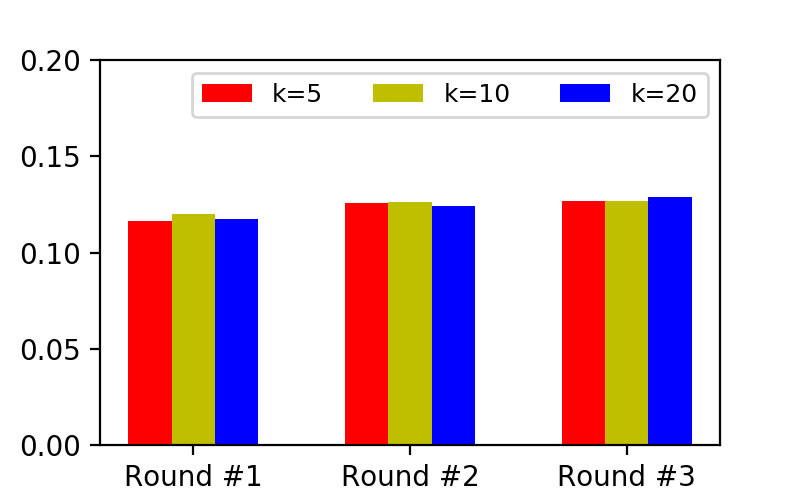} 
    \caption{CCER QS-2} 
  \end{subfigure} 
  \begin{subfigure}[t]{0.24\linewidth}
    \includegraphics[width=1\linewidth]{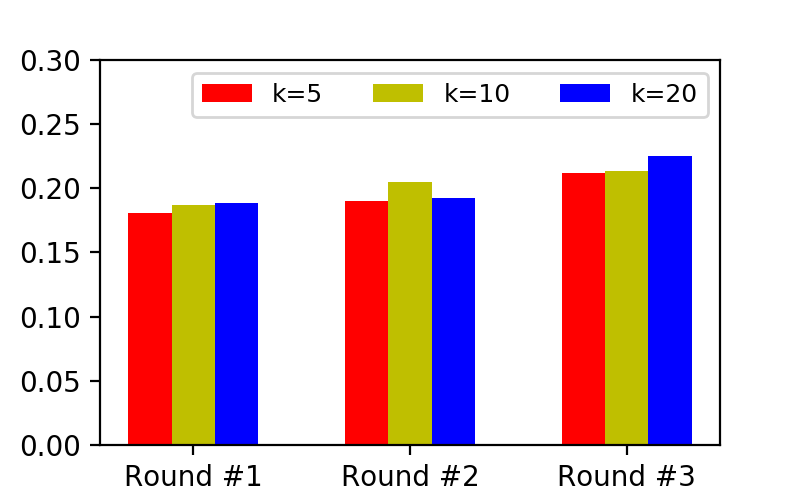} 
    \caption{SD QS-1} 
  \end{subfigure}
  \begin{subfigure}[t]{0.24\linewidth}
    \includegraphics[width=1\linewidth]{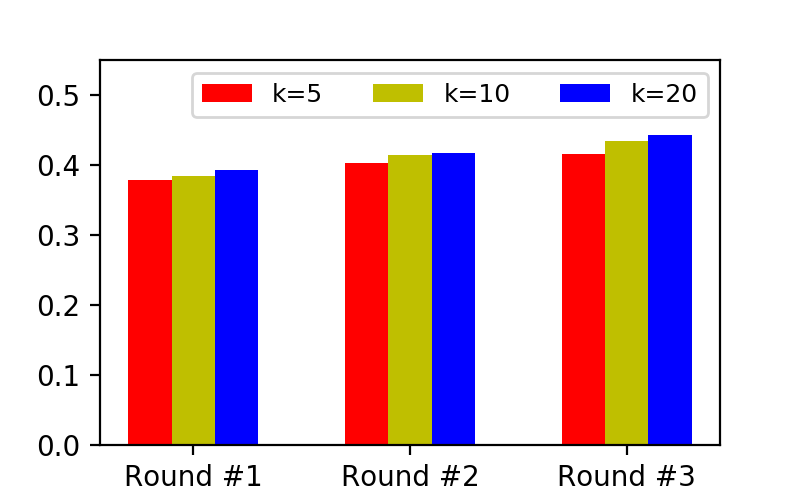} 
    \caption{SD QS-2} 
  \end{subfigure} 
  \captionshrink
  \caption{Impact of the cutoff parameter $k$ for Core Column Entity Ranking (CCER) and Schema Determination (SD).}
  \captionshrink
  \label{fig:iter_k} 
\end{figure*}

Next, we assess the impact of the number of feedback items leveraged, that is, the value of $k$ when using the top-$k$ schema labels in core column entity ranking and top-$k$ entities in schema determination.
Figure~\ref{fig:iter_k} shows how performance changes with different $k$ values.  For brevity, we report only on NDCG@10 and note that a similar trend was observed for NDCG@5.
We find that the differences between the different $k$ values are generally small, with $k=10$ being a good overall choice across the board. 


To further analyze how individual queries are affected over iterations, Table~\ref{tbl:hhs} reports the number of queries that are helped ($\uparrow$), hurt ($\downarrow$), and remained unchanged ($-$).  We define change as a difference of $\geq$0.05 in terms of NDCG@10.
We observe that with the exception of schema determination on QS-2, the number of queries hurt always decreases between successive iterations.  Further, the number of queries helped always increases from Round \#1 to \#3.

\begin{figure}[b]
   \centering
   \includegraphics[width=0.48\textwidth]{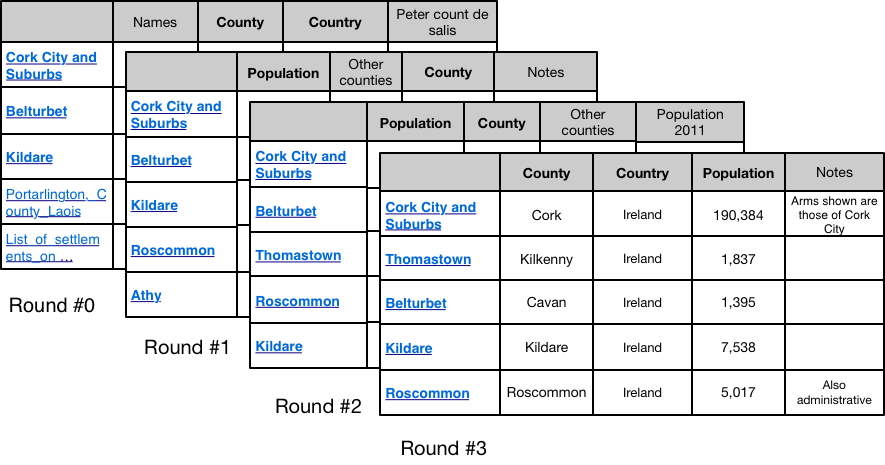} 
   \vspace*{-\baselineskip}
   \caption{Generated table in response to the query ``Towns in the Republic of Ireland in 2006 Census Records.'' Relevant entities and schema labels are boldfaced.}
   \vspace*{-0.5\baselineskip}
\label{fig:re_example}
\end{figure}
\begin{table}[t]
\center
\caption{The number queries helped ($\Delta$NDCG@10$\geq$0.05), hurt ($\Delta$NDCG@10$\leq$-0.05), and unchanged (remaining) for core column entity ranking (CCER) and schema determination (SD).}
  \captionshrink
\label{tbl:hhs}
	\begin{tabular}{lc ccc c ccc}
	\toprule
	&& \multicolumn{3}{c}{\textbf{CCER}} & & \multicolumn{3}{c}{\textbf{SD}} \\
	\cline{3-5} \cline{7-9} 
	QS-1 &&  $\uparrow$ & $\downarrow$ & $-$  && $\uparrow$& $\downarrow$ & $-$  \\
	\midrule
	Round \#0 vs. \#1 && 43 & 38 & 38 && 52 & 7 & 60 \\
	Round \#0 vs. \#2 && 50 & 30 & 39 && 61 & 5 & 53 \\
	Round \#0 vs. \#3 && 49 & 26 & 44 && 59 & 2 & 58 \\

	\midrule
	QS-2 && $\uparrow$ & $\downarrow$ & $-$  && $\uparrow$& $\downarrow$ & $-$  \\
	\midrule
	Round \#0 vs. \#1 && 166 & 82 & 346 && 386 & 56 & 158 \\
	Round \#0 vs. \#2 && 173 & 74 & 347 && 388 & 86 & 126 \\
	Round \#0 vs. \#3 && 173 & 72 & 349 && 403 & 103 & 94 \\

	\bottomrule
	\end{tabular}
\end{table}

Lastly, we demonstrate how results change over the course of iterations, we show one specific example table in Fig.~\ref{fig:re_example} that is generated in response to the query ``Towns in the Republic of Ireland in 2006 Census Records.''

\subsection{Parameter Learning}
\label{sec:anal:param}

For simplicity, we have so far used all features with equal weights for core column entity ranking (cf. Eq.~\ref{eq:ccer}) and schema determination (cf. Eq.~\ref{eq:sd}).  Here, we aim to learn the feature weights from training data.  In Tables~\ref{tbl:er_5fold} and~\ref{tbl:sd_5fold} we report results with weights learned using five-fold cross-validation.  These results are to be compared against the uniform weight settings in Tables~\ref{tbl:er_qs} and~\ref{tbl:sd_qs}, respectively.
We notice that on QS-1, most evaluation scores are lower with learned weights than with uniform weights, for both core column entity ranking and schema determination.  This is due to the fact that queries in this set are very heterogeneous~\citep{Hasibi:2017:DVT}, which makes it difficult to learn weights that perform well across the whole set.
On QS-2, according to expectations, learning the weights can yield up to 18\% and 21\% relative improvement for core column entity ranking and schema determination, respectively.

\begin{table}[t]
\center
\caption{Core column entity retrieval results with parameters learned using five-fold cross-validation. In parentheses are the relative improvements w.r.t. using uniform weights.}
  \captionshrink
	\begin{tabular}{@{~}lll@{~~~}c@{~~~}ll@{~}}
	\toprule
    & \multicolumn{2}{c}{\textbf{QS-1}} && \multicolumn{2}{c}{\textbf{QS-2}} \\
    \cline{2-3} \cline{5-6} 
    \textbf{Method} 
    	& \multicolumn{1}{c}{\textbf{\footnotesize NDCG@5}} 
    	& \multicolumn{1}{c}{\textbf{\footnotesize NDCG@10}} 
    	& & \multicolumn{1}{c}{\textbf{\footnotesize NDCG@5}} 
    	& \multicolumn{1}{c}{\textbf{\footnotesize NDCG@10}}  \\
	\midrule
	Round \#0 & 0.2523 {\footnotesize (-11\%)} & 0.2653 {\footnotesize (-6\%)} && 0.1003 {\footnotesize (+18\%)} & 0.1048 {\footnotesize (+14\%)} \\
	Round \#1  & 0.2782 {\footnotesize (-8\%)} & 0.2772 {\footnotesize (-4\%)} && 0.1308 {\footnotesize (+6\%)} & 0.1252 {\footnotesize (+4\%)}\\
	Round \#2 & 0.3179 {\footnotesize (-6\%)} & 0.3180 {\footnotesize (-1\%)} && 0.1367 {\footnotesize (+5\%)} & 0.1323 {\footnotesize (+5\%)} \\
	Round \#3 & 0.3192 {\footnotesize (-7\%)} & 0.3109 {\footnotesize (-4\%)} && 0.1395 {\footnotesize (+4\%)} & 0.1339 {\footnotesize (+5\%)}\\
	Oracle & 0.3017 {\footnotesize (-14\%)} & 0.3042 {\footnotesize (-9\%)} && 0.1728 {\footnotesize (+9\%)} & 0.1630 {\footnotesize (+5\%)} \\
	\bottomrule	
	\end{tabular}	
	\vspace*{-\baselineskip}	
\label{tbl:er_5fold}
\end{table}
%

\subsection{Feature Importance}

%

To measure the importance of individual features, we use their average learned weights (linear regression coefficients) across all iterations.  The ordering of features for core column entity ranking and QS-1 is: 
$\phi_1 (0.566)$ $>$ $\phi_7 (0.305)$ $>$ $\phi_6 (0.244)$ $>$ $\phi_2 (0.198)$ $>$ $\phi_5 (0.127)$ $>$ $\phi_4 (0.09)$ $>$ $\phi_3 (0.0066)$.
For QS-2 it is: 
$\phi_7 (0.298)$ $>$ $\phi_1 (0.148)$ $>$ $\phi_3 (0.108)$ $>$ $\phi_4 (0.085)$ $>$ $\phi_5 (0.029)$ $>$ $\phi_2 (-0.118)$ $>$ $\phi_6 (-0.128)$.
Overall, we find the term-based matching (Language Modeling) score ($\phi_1$) and our novel entity-schema compatibility score ($\phi_7$) to be the most important features for core column entity ranking.
Turning to schema determination, on QS-1 the ordering is: 
$\phi_5 (0.23)$ $>$ $\phi_3 (0.076)$ $>$ $\phi_1 (-0.035)$ $>$ $\phi_2 (-0.072)$ $>$ $\phi_4 (-0.129)$.  
For QS-2 it is:
$\phi_5 (0.27)$ $>$ $\phi_4 (0.181)$ $>$ $\phi_1 (0.113)$ $>$ $\phi_3 (0.018)$ $>$ $\phi_2 (-0.083)$.
Here, entity-schema compatibility ($\phi_5$) is the single most important feature on both query sets.


\begin{table}[t]
\center
\caption{Schema determination results with parameters learned using five-fold cross-validation. In parentheses are the relative improvements w.r.t. using uniform weights.}
  \captionshrink
	\begin{tabular}{@{~}lll@{~~~}c@{~~~}ll@{~}}
	\toprule
    & \multicolumn{2}{c}{\textbf{QS-1}} && \multicolumn{2}{c}{\textbf{QS-2}} \\
    \cline{2-3} \cline{5-6} 
    \textbf{Method} 
    	& \multicolumn{1}{c}{\textbf{\footnotesize NDCG@5}} 
    	& \multicolumn{1}{c}{\textbf{\footnotesize NDCG@10}} 
    	& & \multicolumn{1}{c}{\textbf{\footnotesize NDCG@5}} 
    	& \multicolumn{1}{c}{\textbf{\footnotesize NDCG@10}}  \\
    \midrule
	Round \#0 & 0.0928 {\footnotesize (+18\%)} & 0.1064 {\footnotesize (+21\%)}  && 0.2326 {\footnotesize (+1\%)} & 0.2710 {\footnotesize (+1\%)} \\
	Round \#1 & 0.1663 {\footnotesize (-1\%)} & 0.2066 {\footnotesize (+11\%)} && 0.3865 {\footnotesize (+16\%)} & 0.4638 {\footnotesize (+12\%)}\\
	Round \#2 & 0.1693 {\footnotesize (-5\%)} & 0.2212 {\footnotesize (+8\%)} && 0.3889 {\footnotesize (+8\%)} & 0.4599 {\footnotesize (+11\%)} \\
	Round \#3 & 0.1713 {\footnotesize (-10\%)} & 0.2321 {\footnotesize (+9\%)} && 0.3915 {\footnotesize (+6\%)} & 0.4620 {\footnotesize (+6\%)}\\
	Oracle & 0.1719 {\footnotesize (-14\%)} & 0.2324 {\footnotesize (-5\%)} && 0.4678 {\footnotesize (+10\%)} & 0.5307 {\footnotesize (+10\%)} \\
	\bottomrule	
	\end{tabular}	
	\vspace*{-0.75\baselineskip}	
\label{tbl:sd_5fold}
\end{table}

\vspace*{-0.5\baselineskip}
\section{Related work}
\label{sec:rw}

Research on web tables has drawn increasing research attention.  We focus on three main related areas: table search, table augmentation, and table mining.

\emph{Table search} refers to the task of returning a ranked list of tables (or tabular data) for a query. Based on the query type, table search can be categorized as keyword-based search~\cite{Cafarella:2008:WEP,Cafarella:2009:DIR,Venetis:2011:RST,Pimplikar:2012:ATQ,Nguyen:2015:RSS} or table-based search~\cite{Lehmberg:2015:MSJ, Ahmadov:2015:THI, DasSarma:2012:FRT, Yakout:2012:IEA, Nguyen:2015:RSS, Limaye:2010:ASW}.  \citet{Zhang:2018:AHT} propose a set of semantic features and fusion-based similarity measures~\citep{Zhang:2017:DPF} for table retrieval with respect to a keyword query. Focusing on result diversity, \citet{Nguyen:2015:RSS} design a goodness measure for table search and selection. There are some existing table search engines, e.g., Google Fusion Tables~\cite{Cafarella:2009:DIR}. Table search is often regarded as a fundamental step in other table related tasks. For example, \citet{DasSarma:2012:FRT} take an input table to search row or column complement tables whose elements can be used for augmenting a table with additional rows or columns. 

\emph{Table augmentation} is about supplementing a table with additional elements, e.g., new columns~\cite{DasSarma:2012:FRT,Cafarella:2009:DIR,Lehmberg:2015:MSJ,Yakout:2012:IEA,Bhagavatula:2013:MEM, Zhang:2018:SAS}. \citet{Zhang:2017:ESA} propose the tasks of row and column population, to augment the core column entity set and column heading labels. They capture relevant data from DBpedia and the WikiTables corpus. Search based on attributes, entities and classes is defined as \emph{relational search}, which can be used for table column augmentation. \citet{Kopliku:2011:TFA} propose a framework to extract and rank attributes from web tables. 
\emph{Data completion} refers to the problem of filling in empty table cells. \citet{Yakout:2012:IEA} address three core tasks: augmentation by attribute name, augmentation by example, and attribute discovery by searching similar tables. Each of these tasks is about extracting table cell data from existing tables. In case that no existing values are captured, \citet{Ahmadov:2015:THI} introduce a method to extract table values from related tables and/or to predict them using machine learning methods.

\emph{Table mining} is to explore and utilize the knowledge contained in tables~\cite{Cafarella:2008:WEP,Sarawagi:2014:OQQ, Yin:2016:NEL, Venetis:2011:RST, Bhagavatula:2015:TEL}. \citet{Munoz:2014:ULD} recover Wikipedia table semantics and store them as RDF triples. A similar approach is taken in \cite{Cafarella:2008:WEP} based on tables extracted from a Google crawl. Instead of mining the entire table corpus, a single table stores many facts, which could be answers to questions. Given a query, \citet{Sun:2016:TCS} identify possible entities using an entity linking method and represent them as a two-node graph question chain, where each node is an entity. Table cells of the KB table are decomposed into relational chains, which are also two-node graphs connecting two entities. The task then boils downing to matching question and table cell graphs using a deep matching model. 
A similar task is addressed by \citet{Yin:2016:NEL} using a full neural network. Information extracted from tables can be used to augment existing knowledge bases~\cite{Sekhavat:2014:KBA, Dong:2014:KVW}. 
Another line of work concerns table annotation and classification. By mining column content, \citet{Zwicklbauer:2013:TDW} propose a method to annotate table headers. Studying a large number of tables in \cite{Crestan:2011:WTC}, a fine-grained table type taxonomy is provided for classifying web tables. 

\section{Conclusion}
We have introduced the task of on-the-fly table generation, which aims to answer queries by automatically compiling a relational table in response to a query.  This problem is decomposed into three specific subtasks: (i) core column entity ranking, (ii) schema determination, and (iii) value lookup.  We have employed a feature-based approach for core column entity ranking and schema determination, combining deep semantic features with task-specific signals.  
We have further shown that these two subtasks are not independent of each other and have developed an iterative algorithm, in which the two reinforce each other. For value lookup, we have entity-oriented fact catalog, which allows for fast and effective lookup from multiple sources.
Using two sets of entity-oriented queries, we have demonstrated the effectiveness of our method. In future work, we wish to consider more heterogeneous table corpus in addition to Wikipedia tables, i.e., arbitrary tables from the Web.


\bibliographystyle{ACM-Reference-Format}
\bibliography{00paper}

\end{document}